\documentclass[conference]{IEEEtran}
\IEEEoverridecommandlockouts
\usepackage{subcaption}
\usepackage{cite}
\usepackage{amsmath,amssymb,amsfonts}

\usepackage[utf8]{inputenc} 
\usepackage[T1]{fontenc}    
\usepackage{hyperref}       
\usepackage{url}            
\usepackage{booktabs}       
\usepackage{amsfonts}       
\usepackage{nicefrac}       
\usepackage{microtype}      
\usepackage{color}
\usepackage{subcaption}
\usepackage{mathtools}
\usepackage{tikz}

\usepackage{epsfig}
\usepackage{graphicx}
\usepackage{amsmath}
\usepackage{amssymb}
\usepackage{amsthm}
\usepackage{verbatim}
\usepackage[numbers]{natbib}

\usepackage{graphicx}
\usepackage{textcomp}
\usepackage{xcolor}

\usepackage{algorithm,algorithmic}
\renewcommand{\algorithmiccomment}[1]{\bgroup\hfill\footnotesize//~#1\egroup}

\DeclareMathOperator*{\argmax}{argmax}

\def\G{\mathbf{G}}
\def\X{\mathbf{X}}

\def\W{\mathbf{W}}
\def\U{\mathbf{U}}

\def\I{\mathbf{I}}

\def\S{\mathbf{S}}

\def\D{\mathbf{D}}

\def\g{\mathbf{g}}

\def\x{\mathbf{x}}

\def\y{\mathbf{y}}

\definecolor{Gray}{gray}{0.85}
\definecolor{LightCyan}{rgb}{0.88,1,1}

\newcommand{\xxx}{\mbox{\textit{FL-MV-DSSM}}}
\newcommand{\xxxx}{\mbox{\textit{FL-DSSM}}}
\newcommand{\xx}{\mbox{\textit{SEMI-FL-MV-DSSM}}}

\begin{document}
	
	\title{A Federated Multi-View Deep Learning Framework for Privacy-Preserving Recommendations}
	

	\author{\IEEEauthorblockN{
		Mingkai Huang\IEEEauthorrefmark{1},
		Hao Li\IEEEauthorrefmark{1},
		Bing Bai\IEEEauthorrefmark{1},
		Chang Wang\IEEEauthorrefmark{1},
		Kun Bai\IEEEauthorrefmark{1},
		Fei Wang\IEEEauthorrefmark{2}
	}
	\IEEEauthorblockA{
		Tencent Inc,
		Cornell University\\
		Email: \IEEEauthorrefmark{1}mingkhuang, leehaoli, icebai, coracwang, kunbai@tencent.com,
		\IEEEauthorrefmark{2}few2001@med.cornell.edu}
	}
	
	\maketitle
	
	\begin{abstract}
		Privacy-preserving recommendations are recently gaining momentum, since the decentralized user data is increasingly harder to collect, by recommendation service providers, due to the serious concerns over user privacy and data security. This situation is further exacerbated by the strict government regulations such as Europe's General Data Privacy Regulations~(GDPR). Federated Learning~(FL) is a newly developed privacy-preserving machine learning paradigm to bridge data repositories without compromising data security and privacy. Thus many federated recommendation~(FedRec) algorithms have been proposed to realize personalized privacy-preserving recommendations. However, existing FedRec algorithms, mostly extended from traditional collaborative filtering~(CF) method, cannot address \emph{cold-start} problem well. In addition, their performance overhead w.r.t. model accuracy, trained in a federated setting, is often non-negligible comparing to centralized recommendations. This paper studies this issue and presents \xxx, a generic content-based federated multi-view recommendation framework that not only addresses the \emph{cold-start} problem, but also significantly boosts the recommendation performance by learning a federated model from multiple data source for capturing richer user-level features. The new federated multi-view setting, proposed by \xxx, opens new usage models and brings in new security challenges to FL in recommendation scenarios. We prove the security guarantees of \xxx, and empirical evaluations on \xxx~and its variations with public datasets demonstrate its effectiveness. Our codes will be released if this paper is accepted.
	\end{abstract}
	
	
	\section{Introduction}
	Privacy-preserving recommendation, motivated by the increasing interest in user privacy, data security, and strict government regulations like Europe's GDPR~\cite{10.5555/3152676}, is recently gaining momentum. Federated Learning~(FL) has been recognized as one of the effective privacy-preserving machine learning paradigms for bridging data repositories while respecting data privacy~\cite{fl_concepts}, thanks to its decentralized collaborative machine learning process without exposing local raw data of any FL participant. Therefore, the combination of recommendation and FL has received widespread attention, which leads to many federated recommendation~(FedRec) algorithms being proposed.
	
	Existing FedRec algorithms are mostly derived from collaborative filtering~(CF)~\cite{usercf, itemcf}, which privately uses user's previous history of interaction to predict the most relevant items for recommendation. The performance overhead of CF-based FedRec~\cite{fcf, fedmf} is observable but acceptable comparing to traditional CF. In addition, inherited from CF, CF-based FedRec suffers from cold-start problem~\cite{cold-start,bai2017dltsr}. The drawback of CF-based FedRec motivates an initial attempt on content-based FedRec, FedNewsRec~\cite{fedrec}, by simply applying FL's FedAvg~\cite{horizon_fl} to a deep learning model designed specifically for news recommendation, which is hard to generalize to other FedRec scenarios. Therefore, the limitations of existing FedRec motivate us to take further steps on addressing both the cold-start problem and recommendation performance. 
	
	To achieve the goals mentioned above, in this work we propose \xxx, a generic content-based federated multi-view recommendation framework. First of all, by transforming a generic deep learning model, Deep Structured Semantic Models~(DSSM~\cite{dssm}) which can map users and items to a shared semantic space for further content-based recommendation, into a federated setting, \xxx~is able to handle existing FedRec's cold-start problem. Then, by designing a novel approach for \xxx~to learn a federated model from multiple data source for capturing richer user-level features, \xxx~greatly boosts its recommendation performance. Moreover, \xxx~presents a new federated multi-view setting, which potentially opens new usage models, e.g. jointly learning a federated model using data from different mobile phone Apps~(e.g. gaming Apps and mobile App markets for accurate mobile game recommendations), and inevitably brings in new challenges, e.g. preventing data leakage among different phone Apps. We will address these challenges in this paper.
	
	The contribution of this paper is threefold. First, to the best of our knowledge, we present the first generic content-based federated multi-view framework and several algorithms that address the cold-start problem and recommendation performance simultaneously. Second, we extend the vanilla FedAvg~\cite{horizon_fl, google_fl_sys_design} from traditional federated setting to a new federated multi-view setting, and correspondingly present a novel approach for securely learning and aggregating multiple local models that share a single model. Third, we carefully study the challenges of the new federated multi-view setting, and present a solution to guarantee its security requirement. Empirical evaluations on \xxx~and its variations with public datasets demonstrate the effectiveness of our framework.
	
	\section{Related Work}
	
	\textbf{Recommendation Systems:}
	In general, traditional recommendation can be divided into CF-based~\cite{usercf, itemcf} recommendation and content-based recommendation~\cite{amazon_rec, rec_click, dssm}. CF relies on user's considerable history data before it can predict the most relevant item for recommendation, and thus suffers from the problem known as cold-start problem~\cite{cold-start}. Content-based recommendation~\cite{amazon_rec, rec_click,dssm,bai2020csrn}, on the other hand, relies on the similarity computed from latent features, learned from user and item information, to recommend items for users, thus it can handle cold-start problem better than CF, but may fail to deliver high quality recommendations when it is difficult to acquire sufficient user information and item information~\cite{mv-dssm,zhang2020general}. In addition, traditional recommendation relies on centralized data for training and prediction.
	
	\textbf{Federated Learning~(FL):}
	FL is a newly developed privacy-preserving machine learning paradigm to bridge data repositories without compromising data security and privacy~\cite{fl_concepts}. By only transmitting summative information~(gradients~\cite{horizon_fl}, dot products~\cite{vertical_fl}, etc.) of training process, rather than collecting user raw data, FL by essence is a distributed decentralized collaborative machine learning framework without exposing FL participant's raw data. In practice, to prevent sensitive information leakage through summative information~\cite{deep_leakage}, FL is further secured by privacy-preserving mechanisms such as DP~\cite{dwork2006calibrating}, HE~\cite{he1978}, MPC~\cite{smc1987}, and so on.
	
	\textbf{Federated Recommendation Algorithms~(FedRec):}
	Similar to traditional recommendation, existing FedRec algorithms can be mainly divided into CF-based, e.g. FCF~\cite{fcf}, FedMF~\cite{fedmf}, etc., and content-based, e.g. FedNewsRec~\cite{fedrec}. Similarly, both FCF and FedMF locally update user matrix on FL participants, and globally aggregate item matrix, through item gradients, on FL server. The difference between FCF and FedMF is that FedMF further protects item gradients with homomorphic encryption to prevent private information leakage through gradients. Similar to traditional CF, both FCF and FedMF cannot handle cold-start problem~\cite{cold-start}. Unlike exiting CF-based FedRec, \xxx~and its variations are all content-based FedRec. FedNewsRec is the first content-based FedRec that uses complicated deep learning models designed specifically towards news recommendation. In addition, the recommendation quality of FedNewsRec mainly depends on its model design. Unlike FedNewsRec, our methods use general DSSM as our basic model to further research on performance optimization methods in federated multi-view setting. 
	
	\textbf{Multi-View Recommendation Algorithms:}
	There are some multi-view recommendation algorithms, e.g. MV-DSSM~\cite{mv-dssm} and FED-MVMF~\cite{fed-mv-cf}. MV-DSSM~\cite{mv-dssm} extends DSSM by utilizing information from different Apps and trains a shared user sub-model, leading to better performance on item recommendation. However, MV-DSSM is originally proposed for multi-item-views learning and requires centralized dataset and thus cannot work in a FL setting. Moreover, our methods consider multi-user-views learning to enrich the user features when training a shared item sub-model. Meanwhile out methods preserve data privacy between user views since we think user sub-model contains user private information that should not be shared across different views. FED-MVMF extends FCF to matrix factorization with multiple data sources, and thus outperforms FCF only using single data source. Comparing to FED-MVMF, our methods are content-based FedRec supporting deep learning model, which is more flexible.
	
	\section{Preliminaries}
	\label{sec:preliminary}
	This section gives our problem definition and reviews related techniques to \xxx. 
	
	\textbf{Problem Definition.}
	We define a federated multi-view setting, the aim of which is to learn a model over data that not only resides on $m$ distributed nodes, but also locates in $n$ isolated views on each distributed node. One possible form of a ``view'' can be a user mobile phone App. As a running example, consider the scenario of recommending a mobile game App in an App market. Not only the user behaviors in App markets~(e.g. Google Play) matter, but also her behaviors in gaming Apps~(e.g. Fortnite) will contribute a lot. However, such data collaboration among multiple business entities is often unachievable due to data security issues if without privacy-preserving FL methods.
	
	We thus denote the decentralized federated multi-view datasets on each distributed node, or a FL client, as ${D}^n = {(\U_1,\I), \ldots, (\U_i,\I), \ldots, (\U_n,\I)}$, where all user view datasets $\U\in \mathbb{R}^{n \times d_{\U}}$ are generated by different views, and item dataset $\I \in \mathbb{R}^{d_{\I}}$ is downloaded from server, e.g. a mobile App's backend service platform. The semantic vectors can be extracted from each view-level user dataset $\U_i \in \mathbb{R}^{d_{\U_i}}$ and item dataset $\I$ respectively by using deep models like DSSM. Then our goal is to find a non-linear mapping $f(\cdot)$ for each view such that the sum of similarities, in the semantic space between mapping of all user view datasets $\U$ and item dataset $\I$, is maximized on each client. Our objective on each client is defined as follows:	
	\begin{align}
	\label{equ:obj}
	\argmax_{\W_\I, \W_{\U_1},\ldots, \W_{\U_n}}\sum_{j = 1}^{S}\frac{\exp(\gamma \cos(\y_\I,\y_{i,j}))}{\sum_{\X^{'}\in \U_i}\exp(\gamma \cos(\y_\I, f_{i}(\X^{'}, \W_{i})))},   
	\end{align}
	where $S$ denote the number of positive user-item pair $(\X_{\U_i,j}, \X_{\I_j})$, $i$ is the index of the view $\U_i$ in sample $j$, $\y$ denote the mapping result of $f(\cdot)$, and $\gamma$ is the temperature parameter. 
	
	\textbf{Security Definition.}
	In addition to the security requirements from traditional FL, \xxx~requires extra security guarantees.  In our federated multi-view setting, although all views collaboratively train a model with datasets $\U$ and $\I$ on each FL client, there should be no raw data interaction between views since each dataset $\U_i$ contains private view-specific information that should be protected. Moreover, each view's contribution to the item sub-model, learned from shared local dataset $\I$, should be protected as well since malicious view could otherwise infer innocent view's raw data from her gradients~\cite{deep_leakage} by monitoring her changes to the shared local item sub-model. 
	
	\textbf{Threat Models.}
	We consider the following threat models in our federated multi-view setting:
	\begin{itemize}
		\item \textbf{[\textit{Traditional FL}]}: FL clients and/or FL server are active adversaries who deviate from the FL protocol, e.g., sending incorrect and/or arbitrarily chosen messages to honest users, aborting, omitting messages, and sharing their entire view of the protocol with each other, and also with the server if server is an active adversary.
		\item \textbf{[\textit{Federated Multi-View}]}: Certain view can be fully malicious, which means as an APP, it would act arbitrarily, e.g. monitoring network interface to observe innocent view's network traffic, making null updates to shared local item sub-model to infer innocent view's update, monitoring changes of item sub-model, etc., in order to infer innocent view's data information. 
	\end{itemize}
	In addition, we make the following assumption:
	\begin{itemize}
		\item \textbf{[\textit{View-Level Isolation}]}: Each view's dataset $\U_i$ and model $\W_{\U_i}$ are only accessible to the $i$-th view, such that malicious view cannot access $\U_i$ and $\W_{\U_i}$. The isolation can be achieved through encryption or TEE~\cite{tee}.
	\end{itemize}
	
	\textbf{Federated Learning.}
	FedAvg~\cite{horizon_fl}, which allows training multiple epochs locally on each FL client before aggregating a global model on FL server,  is widely used in FL setting. The aggregation process is defined as below:
	\begin{equation}
	\W = \sum^K_{k=1}\frac{m_k}{m}\W_k,
	\end{equation}
	where $K$ is the number of clients, $m$ is the number of decentralized samples, and $m_k$ is the number of samples on the $k$-th FL client.
	
	\textbf{Deep Structured Sematic Models~(DSSM).} 
	The DSSM~\cite{dssm}, originally designed for web search, can extract semantic vectors from user's query words and candidate documents, by multi-layer neural networks, and then employ cosine similarity to measure the relevance between query and documents in semantic space. In our generic federated multi-view recommendation setting, we adopt DSSM as our basic model, shown in Figure \ref{pic:dssm}, and extend it into \xxx. Specifically, in \xxx, DSSM's user query is equivalent to \xxx's user feature of $i$-th view $\U_i$, and document is equivalent to item $\I$. 
	
	More formally, if we denote $\x$ as the original feature vector of query words or documents, $\y$ as the semantic vector, $l_i, i = 1, \dots, N - 1$, as the intermediate hidden layers, $\W_i$ as the $i$-th weight matrix, and $b_i$ as the $i$-th bias term, we have DSSM's forward propagation process defined as:
	\begin{equation}
	\label{equ:forward}
	\begin{split}
	l_1 &= \W_1\x, \\
	l_i &= f(\W_il_{i-1} + b_i), i = 2, \dots, N - 1, \\
	\y &= f(\W_Nl_{N-1} + b_N).
	\end{split}
	\end{equation}
	
	The semantic relevance score between a query $Q$ and a document $D$ is then measured as:
	\begin{equation}
	R(Q, D) = cosine(\y_Q, \y_D) = \frac{\y_Q^\top \y_D}{\left \| \y_Q \right \| \cdot \left \| \y_D \right \|},
	\end{equation}
	where $\y_Q$ and $\y_D$ are semantic vectors of query and document, respectively.
	
	We assume that a query is relevant to the documents that are clicked on for that query, and the parameters of the DSSM, i.e., the weight matrix $\W$ are optimized using this information to maximize the conditional likelihood of the clicked documents given queries. The posterior probability of a document given a query is calculated from the semantic relevance score between them through a softmax function
	\begin{equation}
	\mathbb{P}(D|Q) = \frac{\exp(\gamma R(Q,D))}{\sum_{D^{'}\in \D} \exp(\gamma R(Q,D^{'}))},
	\end{equation}
	where $\gamma$ is the temperature parameter in the softmax function. The $\D$ denotes the set of candidate documents to be ranked. In practice, for each pair, $\langle$query, clicked-document$\rangle$, denoted by $(Q,D^{+})$ where $Q$ is a query and $D^{+}$ is the clicked document, we approximate $\D$ by including $D^{+}$ and $N$ randomly selected unclicked documents, denote by $\{D_{j}^{-}; j=1,,N\}$. In training, the loss function we need to minimize is
	\begin{equation}
	L(\Lambda) = -\log \prod_{(Q,D^{+})} P(D^{+}|Q),
	\end{equation}
	where $\Lambda$ denotes the parameter set of the neural networks. 
	
	\section{\xxx: Generic Multi-View Federated Recommendation Framework}
	\label{sec:alg}
	
	\begin{figure*}[ht]
		\begin{subfigure}[c]{.4\textwidth}
			\centering
			\includegraphics[width=\linewidth]{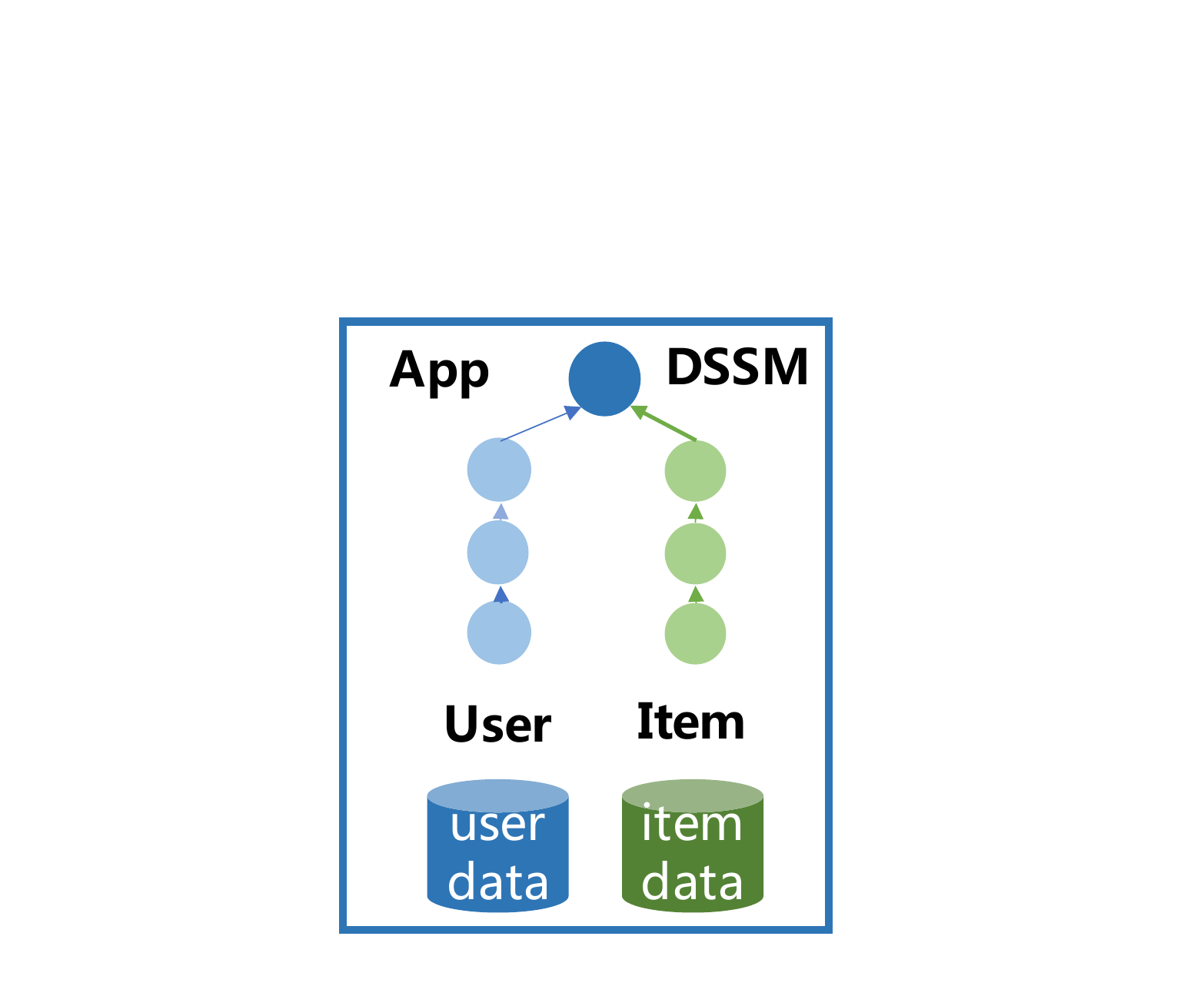}
			\caption{DSSM}
			\label{pic:dssm}
		\end{subfigure}
		\begin{subfigure}[c]{.4\textwidth}
			\centering
			\includegraphics[width=\linewidth]{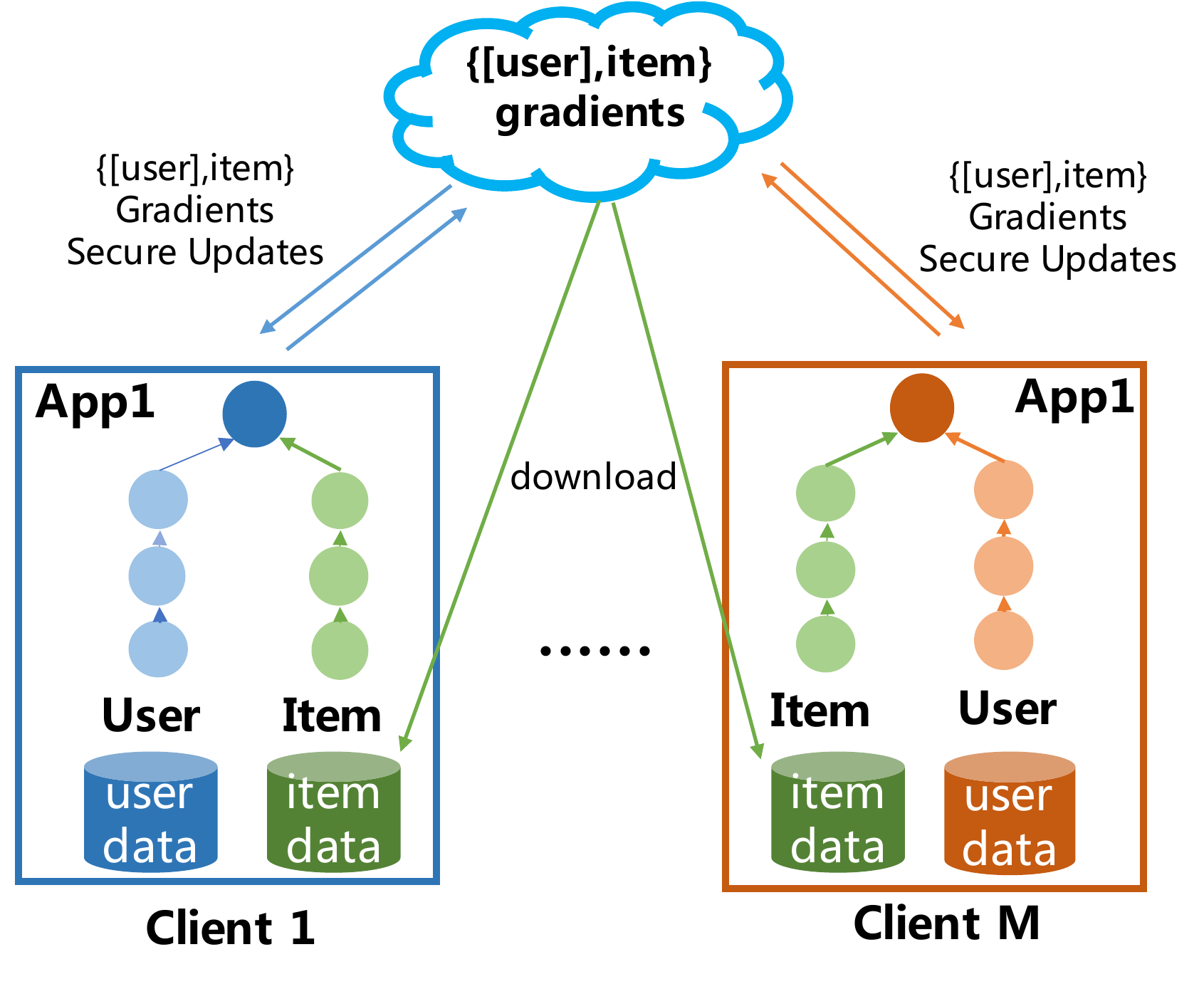}
			\caption{\xxxx}
			\label{pic:fl-dssm}
		\end{subfigure}
		\newline
		\newline
		\begin{subfigure}[c]{.5\textwidth}
			\centering
			\includegraphics[width=\linewidth]{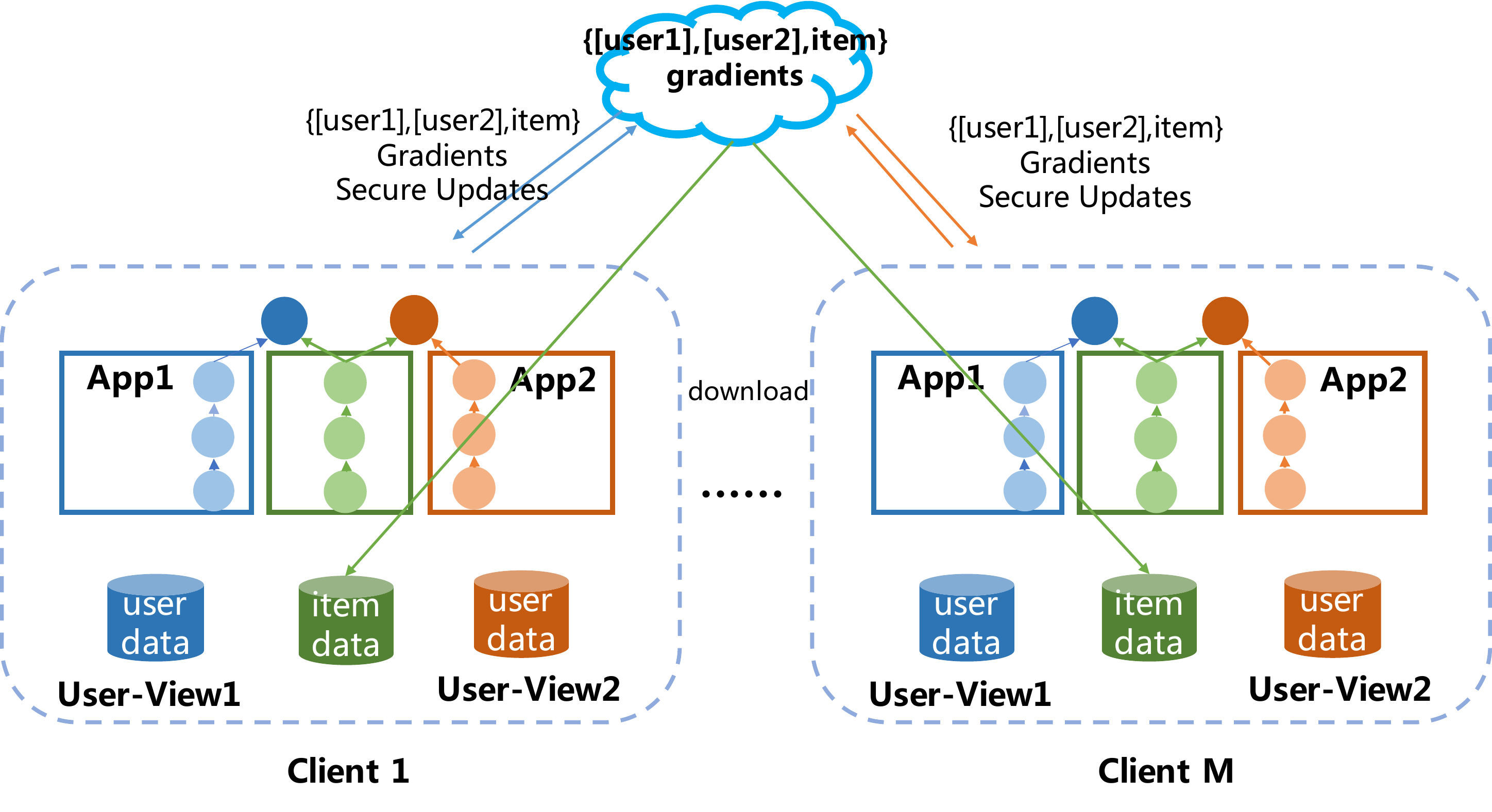}
			\caption{\xxx}
			\label{pic:fl-mv-dssm}
		\end{subfigure}
		\begin{subfigure}[c]{.5\textwidth}
			\centering
			\includegraphics[width=\linewidth]{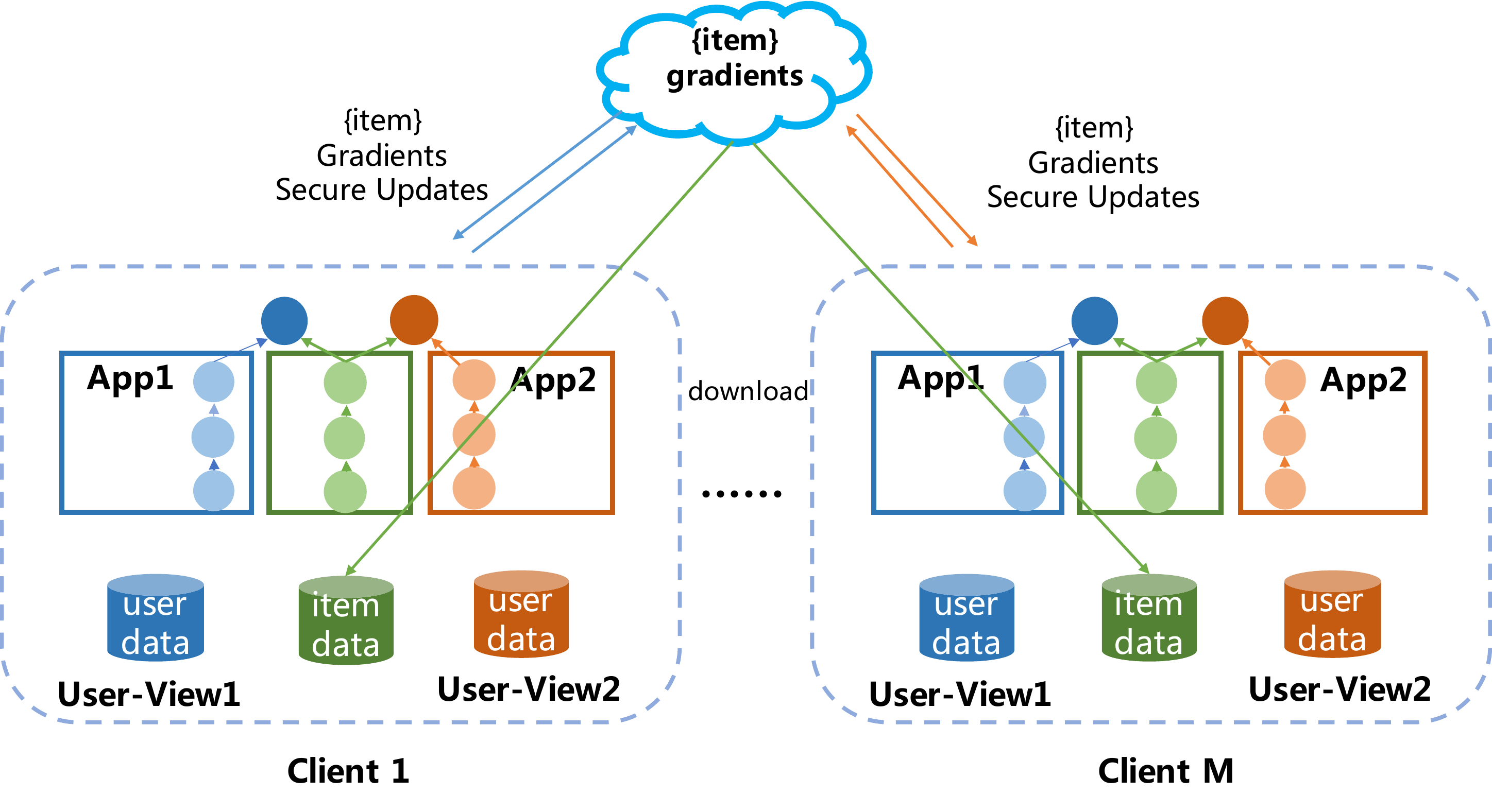}
			\caption{\xx}
			\label{pic:semi-fl-mv-dssm}
		\end{subfigure}
		\caption{\xxx~and its variations.}
		\label{fig:framework}
	\end{figure*}
	This section presents our generic multi-view federated learning framework. Specifically, regarding recommendation scenario, we first introduce \xxx's training and prediction algorithms respectively, and then show how \xxx~guarantees data privacy and user privacy between different views in federated multi-view setting. We also present some variations of \xxx, such as \xxxx~and \xx, as shown in Figure \ref{fig:framework}.
	
	\subsection{\xxx~Training Algorithm}
	We assume each FL client has $N$ views~(Apps) of user-level features, denoted as $\U_i$ for the $i$th view. The $i$th view~(App) can only access the $\U_i$ dataset. The item dataset $\I$ is downloaded from recommendation provider. All views can access the local shared dataset $\I$. Regarding a FedRec task, old users are assumed to have some behavior data that can generate $\y$, while new users don't have any behavior data. \xxx~builds on traditional FedAvg~\cite{horizon_fl} algorithm, which requires FL server to provide initial models.
	
	\begin{algorithm}[htb]
		\caption{\xxx~Training Algorithm}
		\label{alg:fl-mv-dssm}{
			\begin{algorithmic}[1]
				\STATE{\textbf{\textit{FL Client:}}}
				\REQUIRE
				Number of views $N$. Dataset $\D = \{(\X_i, \y), i \in \{1, \ldots, N\}\}$, where $\y$ is user-item behavior data, and $\X_i = (\U_i, \I)$,  where $\U_i$ is the user dataset from view~(e.g. App) $i$,  and $\I$ is the downloaded item dataset to be recommended. Number of FL training round $T$. Learning rate $\eta$.  Initial user sub-models: $\{\W_{\U_1}^0,\ldots,\W_{\U_N}^0\}$. Initial item sub-model weights: $\W_\I^0$. 
				\ENSURE 
				The user sub-models weights: $\{\W_{\U_1}^T,\ldots,\W_{\U_N}^T\}$. The item sub-model weights: $\W_\I^T$.
				\FOR{$k=1:T$}
				\FOR{each view $i=1:N$}
				\STATE{$(\g_\I^k)_{i} = \frac{\partial L(\W_\I^k, \W_{\U_i}^k, \X_i, \y)}{\partial \W_\I^k}$},
				\STATE{$(\g_{\U_i}^k)_{i} = \frac{\partial L(\W_\I^k, \W_{\U_i}^k, \X_i, \y)}{\partial \W_{\U_i}^k}$}.
				\ENDFOR
				\STATE{$\g_\I^k = local\_secure\_aggregate(\{(\g_{\I}^k)_{i}\}, $\\ $i \in \{1, \ldots, N\})$}.
				\STATE{$\G_\I^k = remote\_secure\_aggregate(\g_\I^k)$}.
				\IF{aggregate\_user\_sub-model}
				\FOR{each view $i=1:N$}
				\STATE{$(\G_{\U_i}^k)_i = remote\_secure\_aggregate((\g_{\U_i}^k)_i)$}.
				\ENDFOR
				\ENDIF
				\STATE{$\W_\I^{k+1} = \W_\I^k - \eta \G_\I^k$}.
				\FOR{each view $i=1:N$}
				\STATE{$\W_{\U_i}^{k+1} = \W_{\U_i}^k - \eta \G_{\U_i}^k$}.
				\ENDFOR
				\ENDFOR
				\item[]
				\STATE{\textbf{\textit{FL Server:}}}
				\REQUIRE
				Number of FL clients $M$. Number of FL training round $T$. Client updates $\g_j^k$ in FL round $k$.
				\ENSURE
				Securely aggregate FL client's summative information, e.g. gradients.
				\FOR{$k=1:T$}
				\STATE{$server\_secure\_aggregate(\{\g_j^k\}, $\\ $j \in \S \subseteq \{1, \ldots, M\})$}.
				\ENDFOR
		\end{algorithmic}}
	\end{algorithm}
	
	Algorithm \ref{alg:fl-mv-dssm} shows \xxx's training algorithm. Assuming in \xxx's training phase, all FL clients are old users having behavior data w.r.t. item dataset $\I$ to generate $\y$. Within each view $i$, gradients of user sub-model and item sub-model are calculated based on $i$th view's private user data $\U_i$ and local shared item data $\I$. Although \xxx~is a content-based FedRec, we empirically found that aggregating gradients of item sub-model leads to better recommendation performance, comparing to only aggregating gradients of user sub-model, the result of which is also found in CF-based FedRec~\cite{fcf, fedmf}. Thus in \xxx, gradients of item sub-model will be aggregated in a FL manner, while the aggregation of user gradients is configurable, by ``aggregate\_user\_sub-model'' flag in line 9 of Algorithm \ref{alg:fl-mv-dssm}, which leads to one variations of \xxx, \xx. After each FL training round, both user and item sub-models are updated according to new global gradients distributed by FL server, in a FedAvg manner.
	
	The gradients for both user and item sub-models contain view specific information that should be protected, thus \xxx~provides two secure aggregation primitives, \textit{local\_secure\_aggregate()} and \textit{remote\_secure\_aggregate()}, to secure both local and remote gradients aggregation. We discuss more about both secure aggregation primitives in Section \ref{sec:privacy}.
	
	\subsection{\xxx~Prediction Algorithm}
	Algorithm \ref{alg:fl-mv-dssm-pred} shows \xxx's prediction algorithm. For each item  $\x_{\I_j}$, old or new, item sub-model output its result $\y_{\I_j}$. Meanwhile for user's output, $\y_{\U}$, which is locally secure aggregated from user sub-models in multiple views, is used to compare with all $\y_{\I_j}$s to determine their similarities. Based on the similarity results, \xxx~will output top-$K$ items for the user, old or new. 
	
	\begin{algorithm}[htb]
		\caption{\xxx~Prediction Algorithm}
		\label{alg:fl-mv-dssm-pred}{
			\begin{algorithmic}[1]
				\STATE{\textbf{\textit{FL Client:}}}
				\REQUIRE
				Number of views $N$. Number of items $M$.
				User sub-models: $\{\W_{\U_i}\}, i \in \{1, \ldots, N\}$, where $\W_{\U_i}$ is user sub-model for view $i$. Item sub-model: $\W_\I$. 
				User feature: $\x_{\U_i}$ for $i$th view. Item features: $\x_{\I_j}$ for $j$th item. 
				\ENSURE 
				List of top-$K$ items for recommendation.
				\FOR{each item $j=1:M$}
				\STATE{Compute $\{\y_{\I_j}\}$, where $\y_{\I_j} = f(\W_{\I}, \x_{\I_j})$ according to Eq.\eqref{equ:forward}.}
				\ENDFOR
				\FOR{each item $j=1:M$}
				\FOR{each view $i=1:N$}
				\STATE{Compute $\y_{\U_i} = f(\W_{\U_i}, \x_{\U_i})$ according to Eq.\eqref{equ:forward}.}
				\STATE{$\mathbb{P}(\x_{\I_j}|\x_{\U_i}) = \frac{\exp(\gamma \cos(\y_{\U_i}, \y_{\I_j}))}{\sum_{\y_\I'\in \{\y_{\I_j}\}}\exp(\gamma \cos(\y_{\U_i}, \y_\I'))}$.}
				\ENDFOR
				\STATE{$\mathbb{P}(\x_{\I_j}|\x_\U) = local\_secure\_aggregate($ \\ 
					$\{ \mathbb{P}(\x_{\I_j}|\x_{\U_i})\}, i \in \{1,\ldots,N\})$}.
				\ENDFOR
				\STATE{Select top-$K$ items by sorting $\{\mathbb{P}(\x_{\I_j}|\x_\U)\}$, where $j \in \{1, \ldots, M\}$}.
		\end{algorithmic}}
	\end{algorithm}
	
	\subsection{Secure Primitives for Privacy Protection}
	\label{sec:privacy}
	To defend against the threat models defined in Section \ref{sec:preliminary}, \xxx~mainly adopts two secure primitives, \textit{local\_secure\_aggregate()} and \textit{remote\_secure\_aggregate()}, used in Algorithm \ref{alg:fl-mv-dssm} and \ref{alg:fl-mv-dssm-pred}. 
	
	The purpose of both \textit{local\_secure\_aggregate()} and \textit{remote\_secure\_aggregate()} is to securely aggregate $N$ vectors, locally or remotely, and return the aggregation results, without exposing raw data of each participant to other participants or the curator, either local \xxx~framework or remote FL server. However, different execution environment leads to different implementations for both primitives. 
	
	\subsubsection{local\_secure\_aggregate()}
	It is mainly used on user's mobile phones. It has two usages. First, securely aggregate $N$ gradients of item sub-models in $N$ different views in \xxx's training algorithm; Second, securely aggregate $N$ probabilities, each of which is the possibility that an item is interesting to a user according to a user view $\U_i$,  in \xxx's prediction algorithm. In practice, $N \geqslant 1$. Because $N$ can be 2, and the aggregated gradient will be sent to each view for next round of training, solutions based on Secure Multi-Party Computation~(MPC)~\cite{secure_aggregation} can not be used to implement \textit{local\_secure\_aggregate()} since the aggregation result is accurate and it will expose another view's gradients by subtracting oneself's own gradients from the result gradients. In addition, the computation cost of MPC-based solution is quadratic in FL client~\cite{secure_aggregation}. Therefore we leverage differential privacy~(DP) to realize \textit{local\_secure\_aggregate()}.
	
	Specifically, we apply Gaussian Mechanism~\cite{dwork2014algorithmic} in DP to line 7 in Algorithm \ref{alg:fl-mv-dssm}, by adding noise to each gradients of item sub-model $(\g_\I^k)_i$ in $i$-th view, before aggregation. For Algorithm \ref{alg:fl-mv-dssm-pred}, we add Gaussian noise to each probability before aggregation. Following moments accountant~\cite{abadi2016deep} work, the concrete steps are:
	\begin{itemize}
		\item \textbf{Step1: Random sub-sampling.} For $N$ views in each client, a random subset $B~(|B|_1 \leq N)$ is sampled in each FL round. 
		\item \textbf{Step2: Gradient clipping.} Clip each gradient in $\ell_2$ norm, i.e., the gradient $(\g_\I^k)_i$ is replaced by $(\g_\I^k)_i/\max(1,\frac{\|(\g_\I^k)_i\|_2}{C})$, for a clipping threshold $C$.
		\item \textbf{Step3: Distorting.} A Gaussian Mechanism is used to distort the sum of all updates. Then, we have
		\begin{align}
		\label{equ:dp}
		\widetilde{(\g_\I^k)} = \frac{1}{|B|}(\sum_{i\in B} (\g_\I^k)_i + \mathcal{N}(0,\sigma^2 C^2)),
		\end{align}
		where the value of $\sigma$ satisfies the Theorem 1 in ~\cite{abadi2016deep}. From the Eq.~\ref{equ:dp}, the average distortion is governed by the value of $\sigma$ and $C$, and from the Theorem 1 in ~\cite{abadi2016deep}, we can know that 
		the value of $\sigma$ is inversely proportional to the value of privacy budget $\epsilon$. For example, with $|B|_1 = N$, $\sigma=4$, $\delta=10^{-3}$, and $T=100$, we have $\epsilon \approx 4.33$ using the moments accountant theory~\cite{abadi2016deep}. We can reduce the noise scale by reducing the value of $|B|_1$ and training rounds $T$ while ensuring our model performance.
	\end{itemize}
	
	\subsubsection{remote\_secure\_aggregate()}
	It is used in \xxx's training algorithm to securely aggregate $N~(\g^k_{\U_i})_i$ and $1~(\g_\I^k)$ on each FL client. The secure aggregation is well studied in traditional FL~\cite{secure_aggregation}, thus we follow the methods and proof of the proposed ``Secure Aggregation Protocol'' to realize the \textit{remote\_secure\_aggregate()} in \xxx. In this way, we can ensure that FL server will only see aggregated result without knowing each FL client's update. In addition, the ``Secure Aggregation Protocol'' protocol can handle client drop-out problem well.
	
	\subsection{\xxx~Variations}
	\xxx~has several variations. 
	
	\textbf{\xxxx.} 
	Based on the \xxx~algorithms introduced in previous sections, both \xxxx~training and prediction algorithms can be simply derived by setting the number of views $N$ to $1$. 
	
	\textbf{\xx.}
	We can get \xx, by setting the ``aggregate\_user\_sub-model'' flag in Algorithm \ref{alg:fl-mv-dssm} to be $false$. \xx~only conducts secure aggregation on gradients of item sub-model, rather than aggregating gradients of user sub-model. 
	
	We will evaluate more on these \xxx~variations in Section \ref{sec:exp}.
	
	\section{Experiments}
	\label{sec:exp}
	This section evaluates \xxx. We have proved in Section \ref{sec:privacy} that \xxx~is able to protect data privacy among different views. Now we mainly address the following questions: (\textbf{Q1})~Is \xxx~able to address cold-start problem? (\textbf{Q2})~How is recommendation performance of \xxx, and its variations?
	
	\subsection{Environmental Setup}
	\label{sec:env}
	We implement \xxx~and its variations in Google's TensorFlow Federated~(TFF) simulation framework~\cite{tff}. For \xxx, we take two user views as an example. Table \ref{tab:struct} shows the DSSM model architectures we use for evaluations.
	
	\begin{table}
		\centering
		\renewcommand{\arraystretch}{1.2}
		\caption{Structures of item sub-model and user sub-models of two views.}\label{tab:struct}
		\begin{tabular}{lccccc}
			\toprule
			& Item & User~(View-1) & User~(View-2)\\
			\midrule
			\mbox{Input} & Input~(4739) & Input~(23)  & Input~(30) \\
			\mbox{Layer1} & Dense~(64) & Dense~(64) & Dense~(64) \\
			\mbox{Layer2} & Dense~(32) & Dense~(32) & Dense~(32) \\
			\mbox{Layer3} & Dense~(16) & Dense~(16) & Dense~(16) \\
			\bottomrule
		\end{tabular}
	\end{table}
	
	\textbf{Data Pre-Processing.}
	Similar to existing FedRec algorithms, we use the popular public dataset, MovieLens-100K~\cite{harper2015movielens}, for our evaluations. The MovieLens-100K not only consists of 100K ratings from 943 users on 1682 items~(movies), but also contains user and item information, e.g. user's age and movie's title. 
	For label pre-processing, we create implicit feedbacks as 1 for all $\langle$user, item$\rangle$ pairs where a user explicitly interacted with an item in the dataset, and 0 for the rest. 
	For user features pre-processing, we randomly sample a portion of MovieLens data and select age (normalized to less than or equal to 1), gender (binary feature), and occupation (one-hot vector) as user features for one view (\textbf{View-1}); meanwhile we use user embedding learned by singular value decomposition~(SVD) from MovieLens' interaction matrix, orthogonal to the data of View-1, as user features for another view (\textbf{View-2}). 
	For item feature pre-processing, we select title and genre, coded with 3-gram representation and a series of bits, respectively.
	
	\textbf{Evaluation Metrics.}
	We consider the following evaluation metrics, including Precision@10, Recall@10, NDCG@10, and AUC. Among the metrics, Precision@10, Recall@10 and NDCG@10 only concentrate on the very top of recommendation list, while AUC evaluates the overall accuracy of recommendations.
	
	
	
	
	\textbf{Hyper-parameter setting.}
	Adam optimizer with learning rate 0.001 is used in centralized training and for server side aggregation of FL setting. SGD optimizer with learning rate 0.2 is used for client side training of FL setting. Batch size is set to 20.
	The dimension is set to 30 for matrix factorization.
	For DP parameters, $C = 0.5, \sigma = 1, \delta = 0.001, |B| = N$.
	
	\subsection{Cold-Start Recommendations}
	\begin{table*}
		\centering
		\renewcommand{\arraystretch}{1.3}
		\caption{Cold-Start recommendation performance of \xxx~on MovieLens dataset with three scenarios. The values denote the mean $\pm$ standard deviation across 3 different model builds.}\label{tab:cold-start}
		\begin{tabular}{lccccc}
			\toprule
			& Precision@10 & Recall@10 & NDCG@10 & AUC \\
			\midrule
			\mbox{CS-Users}  & $0.4574\pm0.0085$ & $0.0696\pm0.0100$  & $0.3697\pm0.0031$ & $0.7793\pm0.0617$ \\
			\mbox{CS-Items}  & $0.1586\pm0.0002$ & $0.1086\pm0.0015$ & $0.1318\pm0.0008$ & $0.5809\pm0.0721$ \\
			\mbox{CS-Users-Items} & $0.1408\pm0.0033$ & $0.1335\pm0.0315$ & $0.1232\pm0.0024$ & $0.5672\pm0.0043$\\
			\bottomrule
		\end{tabular}
	\end{table*}
	
	To evaluate cold-start recommendations, we mainly focus on \xxx, and conduct three cold-start scenarios: cold-start users~(CS-Users), cold-start items~(CS-Items), and cold-start users-items~(CS-Users-Items). For the case of cold-start users, a random subset of 10\% users and their interaction data are completely excluded during model training and model parameters are learned with the remaining 90\% of the users and their interaction data. For the case of cold-start items, a random subset of 10\% items are left-out during model training and model parameters are learned with the remaining 90\% of the items. For the case of cold-start users-items, a random subset of 10\% users and items are excluded from the model training and model parameters are learned with the rest of users, interaction data, and items. We use the 10\% held-out datasets in all three scenarios as our testing datasets. 
	
	Table \ref{tab:cold-start} shows the results of three cold-start recommendations. The results demonstrate that without loss of generality, \xxx~can be used for cold-start recommendation reliably. Specifically, the results also show that \xxx~achieves good cold-start prediction performance for a new user, which is valuable for privacy-preserving recommendations since new users are continuously enrolled in the recommendation service. However, the performance of cold-start items and users-items are lower than that of cold-start users. The reason behind this might be that in our datasets, the difference between users (considering age, gender, occupation, etc.) is less than that of items (movie title, genres, etc.), and \xxx~could learn this difference correctly and recommend with higher precision.
	
	\subsection{Recommendation Performance}
	\begin{table*}
		\centering
		\renewcommand{\arraystretch}{1.3}
		\caption{Recommendation performance of \xxx, its variations, and existing FedRec algorithms on MovieLens dataset, after 100 FL training rounds. The values denote the mean $\pm$ standard deviation across 3 different model builds.}\label{tab:perf}
		\begin{tabular}{lccccc}
			\toprule
			& Precision@10 & Recall@10 & NDCG@10 & AUC \\
			\midrule
			\mbox{Centralize-DSSM(View-1)}  & $0.2804\pm0.0011$ & $0.0870\pm0.0036$ & $0.2389\pm0.0013$ & $0.8479\pm0.0009$ \\
			\mbox{Centralize-DSSM(View-2)}  & $0.3980\pm0.0025$ & $0.1501\pm0.0009$ & $0.3401\pm0.0035$ & $0.9202\pm0.0011$ \\
			\mbox{FCF~\cite{fcf}}  & $0.3010\pm0.0032$ & $0.1017\pm0.0019$ & $0.2606\pm0.0008$ & $0.8627\pm0.0003$ \\
			\mbox{FED-MVMF~\cite{fed-mv-cf}}  & $0.3223\pm0.0017$ & $0.1193\pm0.0041$ & $0.2797\pm0.0038$ & $0.8833\pm0.0017$ \\
			\mbox{\xxxx(View-1)}  & $0.2691\pm0.0015$ & $0.0721\pm0.0027$ & $0.2292\pm0.0018$ & $0.8445\pm0.0523$ \\
			\mbox{\xxxx(View-2)}  
			& $0.2656\pm0.0021$ & $0.0676\pm0.0137$  & $0.2186\pm0.0023$ & $0.8398\pm0.0314$ \\
			\mbox{\xxx} & $0.2845\pm0.0034$ & $0.0805\pm0.0113$ & $0.2369\pm0.0104$ & $0.8490\pm0.0033$\\
			\mbox{\xx} & $0.3512\pm0.0070$ & $0.1217\pm0.0214$ & $0.3136\pm0.0038$ & $0.8986\pm0.0051$\\
			\bottomrule
		\end{tabular}
	\end{table*}
	
	To evaluate recommendation performance, we examine \xxx, \xxxx, \xx, centralized DSSM, and existing FedRec algorithms including FCF\cite{fcf} and FED-MVMF\cite{fed-mv-cf}. Specifically, for \xxx~related evaluations, we use two generated user views, View-1 and View-2 as described in Section \ref{sec:env}, to train two user sub-models. For \xxx~and \xx, we randomly select 100 users within each FL training round, and for each user, the two sub-models share a single item sub-model, as introduced in Figure\ref{pic:fl-mv-dssm} and Figure \ref{pic:semi-fl-mv-dssm}, to form a \xxx~task. For \xxxx, we launch two FL training tasks, each of which randomly selects 100 users within each FL training round, and each user sub-model is paired with a item sub-model. For centralized DSSM, the datasets for all users are centralized and trained at the single place. For both FCF and FED-MVMF, we follow the experimental setups in their papers. All the datasets, centralized or decentralized, are randomly divided into an 80\% training set and 20\% testing set.
	
\begin{figure*}[ht]
		\begin{subfigure}{.5\textwidth}
			\label{fig:perf1}
			\centering
			\includegraphics[width=\linewidth]{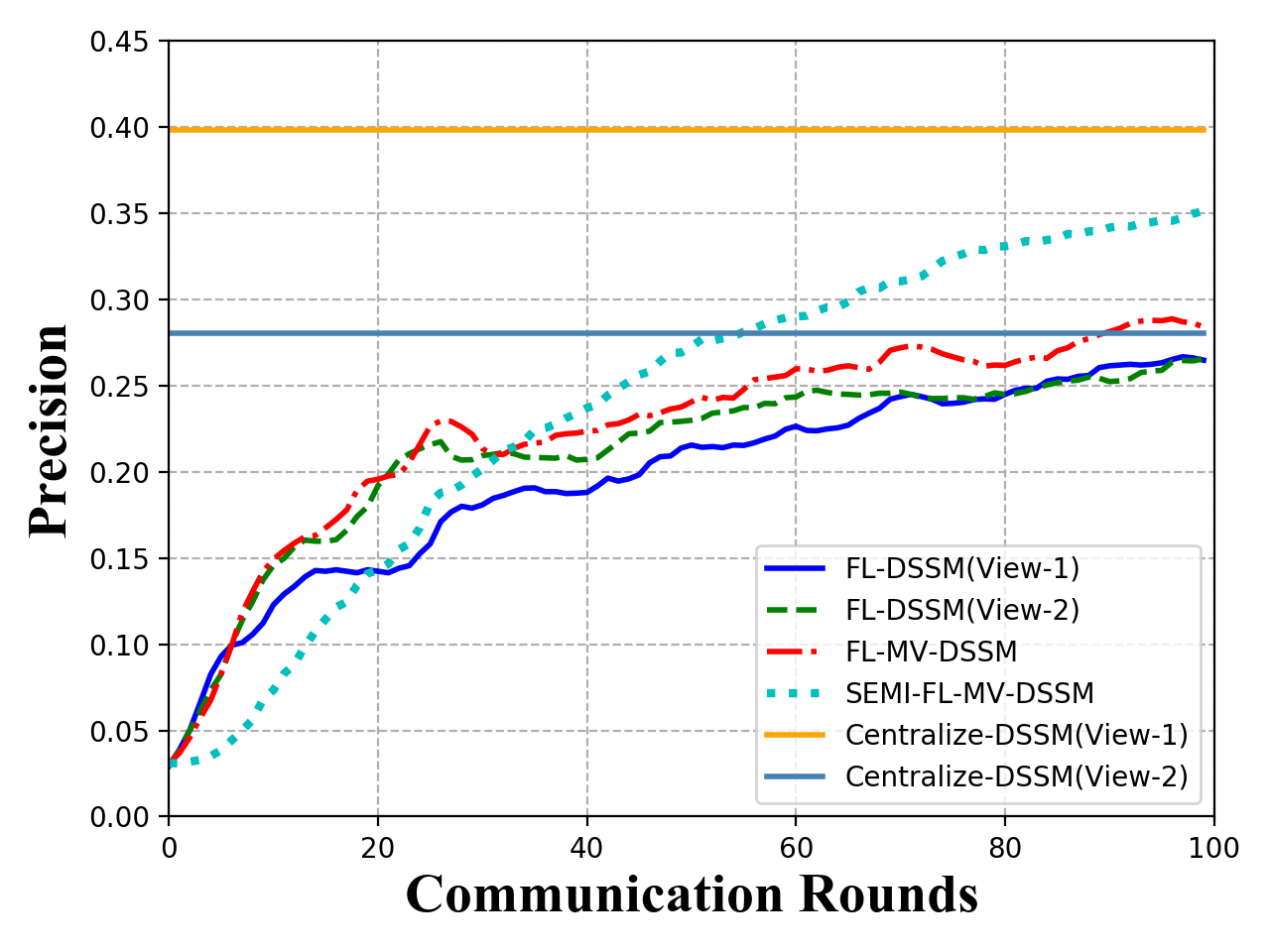}
			\caption{Precision}
		\end{subfigure}
		\begin{subfigure}{.5\textwidth}
			\label{fig:perf2}
			\centering
			\includegraphics[width=\linewidth]{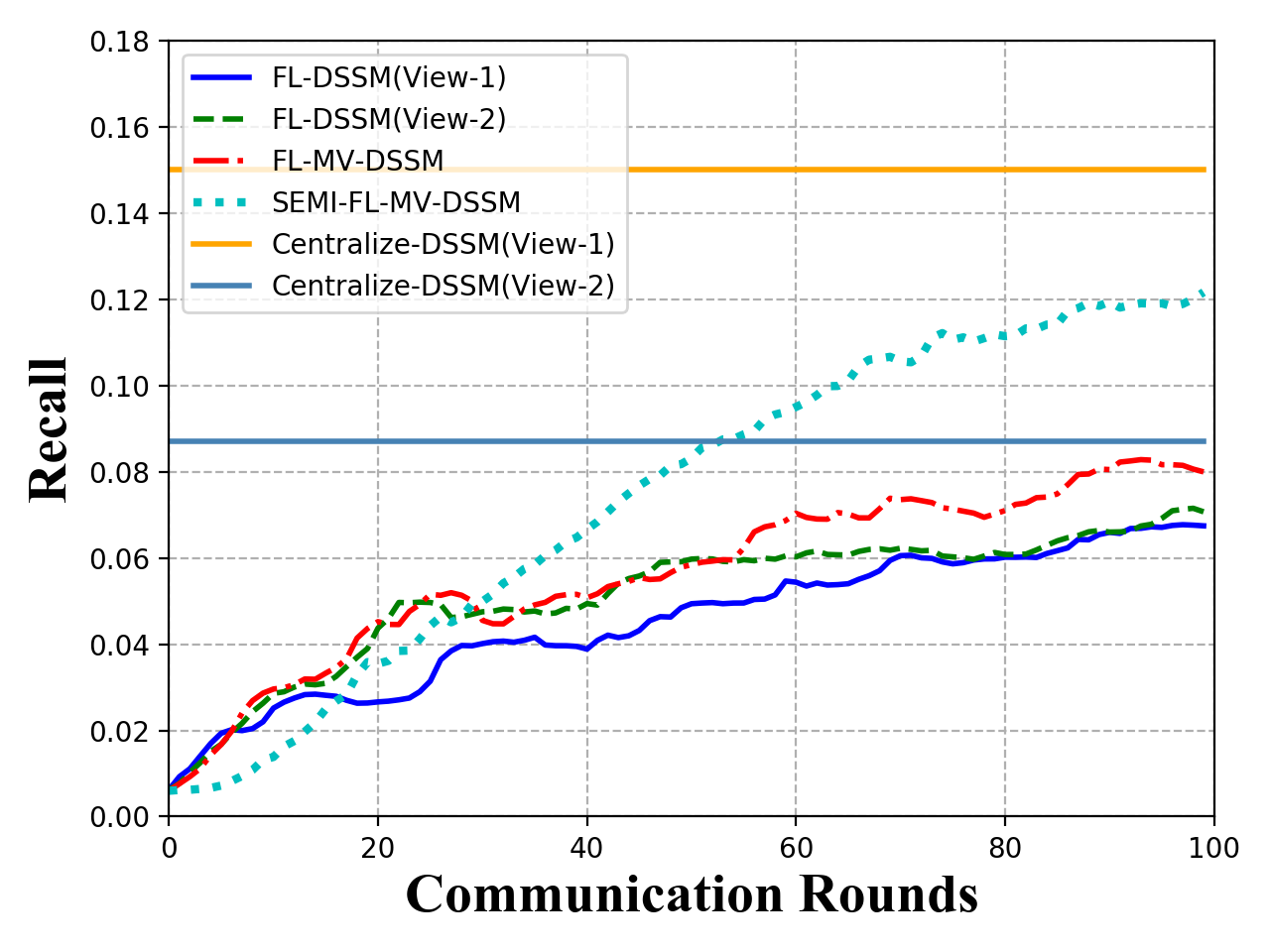}
			\caption{Recall}
		\end{subfigure}
		\caption{Recommendation Performance of \xxx~and its variations.}
		\label{fig:perf}
	\end{figure*}

	Figure \ref{fig:perf} shows the recommendation performance, during the FL training process, precision and recall, of \xxx, \xx, \xxxx, and centralized DSSM. Final results about \xxx~and its related works are listed in Table \ref{tab:perf}. From the results we can see that among FedRec algorithms, \xxx~achieves better performance than \xxxx, since \xxx~can incorporate more user features from multiple views, e.g. from multiple user Apps, to jointly train a better model. One interesting result is that we find \xx, which is only aggregating the shared item sub-model but not the user sub-models, achieves the best performance among FedRec algorithms including FCF and FED-MVMF, and its result is even better than one of the centralized DSSM's result after 60 FL training rounds. This is understandable in that for all FedRec algorithms, their performance data are collected through ``federated evaluation~\cite{leaf}'', and the performance of user sub-model will fit to user local data fast if not aggregating the contributions from other FL participants. We also observe some variations of \xx's performance, caused by randomness in user selection.
	
	\section{Conclusions}
	
	This paper presents \xxx, the first generic content-based federated multi-view framework that could address cold-start problem and recommendation quality at the same time. In addition, this paper extends the traditional federated setting into a new federated multi-view setting, which might potentially enable new usage models of FL in recommendation scenario and bring in new security challenges. By carefully studying the challenges, this paper presents a novel solution addressing the security requirements. Thorough empirical evaluations on \xxx~and its variations with public datasets demonstrate that \xxx~can address cold-start problem, and boost recommendation performance significantly.	
	
	\bibliographystyle{plain}
	\bibliography{biblio}

\begin{thebibliography}{10}

\bibitem{abadi2016deep}
Martin Abadi, Andy Chu, Ian Goodfellow, H~Brendan McMahan, Ilya Mironov, Kunal
  Talwar, and Li~Zhang.
\newblock Deep learning with differential privacy.
\newblock In {\em Proceedings of the 2016 ACM SIGSAC Conference on Computer and
  Communications Security}, pages 308--318, 2016.

\bibitem{fcf}
Muhammad Ammad{-}ud{-}din, Elena Ivannikova, Suleiman~A. Khan, Were Oyomno,
  Qiang Fu, Kuan~Eeik Tan, and Adrian Flanagan.
\newblock Federated collaborative filtering for privacy-preserving personalized
  recommendation system.
\newblock {\em CoRR}, abs/1901.09888, 2019.

\bibitem{usercf}
Robert~M. Bell and Yehuda Koren.
\newblock Improved neighborhood-based collaborative filtering.
\newblock 2007.

\bibitem{google_fl_sys_design}
Keith Bonawitz, Hubert Eichner, Wolfgang Grieskamp, Dzmitry Huba, Alex
  Ingerman, Vladimir Ivanov, Chloe~M Kiddon, Jakub Konecny, Stefano
  Mazzocchi, Brendan McMahan, Timon~Van Overveldt, David Petrou, Daniel Ramage,
  and Jason Roselander.
\newblock Towards federated learning at scale: System design.
\newblock In {\em SysML 2019}, 2019.
\newblock To appear.

\bibitem{secure_aggregation}
Keith Bonawitz, Vladimir Ivanov, Ben Kreuter, Antonio Marcedone, H.~Brendan
  McMahan, Sarvar Patel, Daniel Ramage, Aaron Segal, and Karn Seth.
\newblock Practical secure aggregation for privacy-preserving machine learning.
\newblock In {\em Proceedings of the 2017 ACM SIGSAC Conference on Computer and
  Communications Security}, CCS ’17, page 1175–1191, New York, NY, USA,
  2017. Association for Computing Machinery.

\bibitem{leaf}
Sebastian Caldas, Peter Wu, Tian Li, Jakub Konecn{\'{y}}, H.~Brendan McMahan,
  Virginia Smith, and Ameet Talwalkar.
\newblock {LEAF:} {A} benchmark for federated settings.
\newblock {\em CoRR}, abs/1812.01097, 2018.

\bibitem{fedmf}
Di~Chai, Leye Wang, Kai Chen, and Qiang Yang.
\newblock Secure federated matrix factorization.
\newblock {\em CoRR}, abs/1906.05108, 2019.

\bibitem{dwork2006calibrating}
Cynthia Dwork, Frank McSherry, Kobbi Nissim, and Adam Smith.
\newblock Calibrating noise to sensitivity in private data analysis.
\newblock In {\em Theory of Cryptography Conference}, pages 265--284. Springer,
  2006.

\bibitem{dwork2014algorithmic}
Cynthia Dwork, Aaron Roth, et~al.
\newblock The algorithmic foundations of differential privacy.
\newblock {\em Foundations and Trends{\textregistered} in Theoretical Computer
  Science}, 9(3--4):211--407, 2014.

\bibitem{mv-dssm}
Ali~Mamdouh Elkahky, Yang Song, and Xiaodong He.
\newblock A multi-view deep learning approach for cross domain user modeling in
  recommendation systems.
\newblock In {\em Proceedings of the 24th International Conference on World
  Wide Web}, WWW ’15, page 278–288, Republic and Canton of Geneva, CHE,
  2015. International World Wide Web Conferences Steering Committee.

\bibitem{fed-mv-cf}
Adrian Flanagan, Were Oyomno, Alexander Grigorievskiy, Kuan~Eeik Tan,
  Suleiman~A Khan, and Muhammad Ammad-Ud-Din.
\newblock Federated multi-view matrix factorization for personalized
  recommendations.
\newblock {\em arXiv preprint arXiv:2004.04256}, 2020.

\bibitem{smc1987}
O.~Goldreich, S.~Micali, and A.~Wigderson.
\newblock How to play any mental game.
\newblock In {\em Proceedings of the Nineteenth Annual ACM Symposium on Theory
  of Computing}, STOC ’87, page 218–229, New York, NY, USA, 1987.
  Association for Computing Machinery.

\bibitem{vertical_fl}
Stephen Hardy, Wilko Henecka, Hamish Ivey{-}Law, Richard Nock, Giorgio Patrini,
  Guillaume Smith, and Brian Thorne.
\newblock Private federated learning on vertically partitioned data via entity
  resolution and additively homomorphic encryption.
\newblock {\em CoRR}, abs/1711.10677, 2017.

\bibitem{harper2015movielens}
F~Maxwell Harper and Joseph~A Konstan.
\newblock The movielens datasets: History and context.
\newblock {\em Acm transactions on interactive intelligent systems (tiis)},
  5(4):1--19, 2015.

\bibitem{dssm}
Po-Sen Huang, Xiaodong He, Jianfeng Gao, Li~Deng, Alex Acero, and Larry Heck.
\newblock Learning deep structured semantic models for web search using
  clickthrough data.
\newblock In {\em Proceedings of the 22nd ACM International Conference on
  Information \& Knowledge Management}, CIKM ’13, page 2333–2338, New York,
  NY, USA, 2013. Association for Computing Machinery.

\bibitem{tff}
Google Inc.
\newblock {TensorFlow Federated Framework}.
\newblock \url{https://www.tensorflow.org/federated}, 2019.

\bibitem{amazon_rec}
G.~{Linden}, B.~{Smith}, and J.~{York}.
\newblock Amazon.com recommendations: item-to-item collaborative filtering.
\newblock {\em IEEE Internet Computing}, 7(1):76--80, 2003.

\bibitem{rec_click}
Jiahui Liu, Peter Dolan, and Elin~R\o{}nby Pedersen.
\newblock Personalized news recommendation based on click behavior.
\newblock In {\em Proceedings of the 15th International Conference on
  Intelligent User Interfaces}, IUI ’10, page 31–40, New York, NY, USA,
  2010. Association for Computing Machinery.

\bibitem{horizon_fl}
Brendan McMahan, Eider Moore, Daniel Ramage, Seth Hampson, and
  Blaise~Ag{\"{u}}era y~Arcas.
\newblock Communication-efficient learning of deep networks from decentralized
  data.
\newblock In Aarti Singh and Xiaojin~(Jerry) Zhu, editors, {\em Proceedings of
  the 20th International Conference on Artificial Intelligence and Statistics,
  {AISTATS} 2017, 20-22 April 2017, Fort Lauderdale, FL, {USA}}, volume~54 of
  {\em Proceedings of Machine Learning Research}, pages 1273--1282. {PMLR},
  2017.

\bibitem{tee}
Sandro Pinto and Nuno Santos.
\newblock Demystifying arm trustzone: A comprehensive survey.
\newblock {\em ACM Comput. Surv.}, 51(6), January 2019.

\bibitem{fedrec}
Tao Qi, Fangzhao Wu, Chuhan Wu, Yongfeng Huang, and Xing Xie.
\newblock Fedrec: Privacy-preserving news recommendation with federated
  learning.
\newblock {\em arXiv preprint arXiv:2003.09592}, 2020.

\bibitem{he1978}
R~L Rivest, L~Adleman, and M~L Dertouzos.
\newblock On data banks and privacy homomorphisms.
\newblock {\em Foundations of Secure Computation, Academia Press}, pages
  169--179, 1978.

\bibitem{itemcf}
Badrul Sarwar, George Karypis, Joseph Konstan, and John Riedl.
\newblock Item-based collaborative filtering recommendation algorithms.
\newblock In {\em Proceedings of the 10th International Conference on World
  Wide Web}, WWW ’01, page 285–295, New York, NY, USA, 2001. Association
  for Computing Machinery.

\bibitem{cold-start}
Andrew~I. Schein, Alexandrin Popescul, Lyle~H. Ungar, and David~M. Pennock.
\newblock Methods and metrics for cold-start recommendations.
\newblock In {\em Proceedings of the 25th Annual International ACM SIGIR
  Conference on Research and Development in Information Retrieval}, SIGIR
  ’02, page 253–260, New York, NY, USA, 2002. Association for Computing
  Machinery.

\bibitem{10.5555/3152676}
Paul Voigt and Axel von~dem Bussche.
\newblock {\em The EU General Data Protection Regulation (GDPR): A Practical
  Guide}.
\newblock Springer Publishing Company, Incorporated, 1st edition, 2017.

\bibitem{fl_concepts}
Qiang Yang, Yang Liu, Tianjian Chen, and Yongxin Tong.
\newblock Federated machine learning: Concept and applications.
\newblock {\em ACM Trans. Intell. Syst. Technol.}, 10(2), January 2019.

\bibitem{deep_leakage}
Ligeng Zhu, Zhijian Liu, and Song Han.
\newblock Deep leakage from gradients.
\newblock In H.~Wallach, H.~Larochelle, A.~Beygelzimer, F.~d\textquotesingle
  Alch\'{e}-Buc, E.~Fox, and R.~Garnett, editors, {\em Advances in Neural
  Information Processing Systems 32}, pages 14774--14784. Curran Associates,
  Inc., 2019.

\end{thebibliography}

\end{document}